\documentclass[fleqn,usenatbib]{mnras}

\usepackage{newtxtext,newtxmath}
\usepackage[T1]{fontenc}
\usepackage{ae,aecompl}

\usepackage{graphicx}	% Including figure files
\usepackage{amsmath}	% Advanced maths commands
\title[Radar observations of Draconid outbursts]{Radar observations of Draconid outbursts}

\author[M. D. Campbell-Brown \& G. Stober]{
M. D. Campbell-Brown,$^{1,2}$\thanks{E-mail: margaret.campbell@uwo.ca}
G. Stober,$^{3}$
C. Jacobi,$^{4}$
J. Kero,$^{5}$
A. Kozlovsky,$^{6}$
M. Lester,$^{7}$
\\
$^{1}$Department of Physics and Astronomy, University of Western Ontario, London, Ontario, N6A 3K7, Canada\\
$^{2}$Institute for Earth and Space Exploration, Western University, London, Ontario, N6A 5B7, Canada\\
$^{3}$Institute of Applied Physics \& Oeschger Center for Climate Change Research, Microwave Physics, University of Bern, Bern,Switzerland\\
$^{4}$Institute for Meteorology, Universität Leipzig, Leipzig, Germany\\
$^{5}$Swedish Institute of Space Physics (IRF), Kiruna, Sweden\\
$^{6}$Sodankyla Geophysical Observatory, University of Oulu, Finland\\
$^{7}$University of Leicester, Leicester, UK\\
}

\date{Accepted XXX. Received YYY; in original form ZZZ}

\pubyear{2021}

\begin{document}
\label{firstpage}
\pagerange{\pageref{firstpage}--\pageref{lastpage}}
\maketitle

\begin{abstract}

The Draconid meteor shower shows strong bursts of activity at irregular intervals, with nearly no activity in intervening years. Five outbursts of the Draconid meteor shower were observed with specular meteor radars in Canada and Europe between 1999 and 2018. The outbursts generally lasted between 6 and 8 hours, and most were not fully visible at a single geographical site, emphasizing the need for observations at multiple longitudes for short-duration shower outbursts. There is at least a factor of two difference in the peak flux as measured on different radars; the initial trail radius effect is undercorrected for Draconid meteors, which are known to be fragile.

\end{abstract}


%%%%%%%%%%%%%%%%%%%%%%%%%%%%%%%%%%%%%%%%%%%%%%%%%%

%%%%%%%%%%%%%%%%% BODY OF PAPER %%%%%%%%%%%%%%%%%%

\section{Introduction}

The Draconid meteor shower (009 DRA) (formerly known as the Giacobinids, after their parent comet 21P/Giacobini-Zinner) is a low-activity shower which has irregular, and sometimes spectacular, outbursts. These outbursts can be rich in bright meteors, or confined to very faint meteors mainly visible to radar \citep{egal2019}. The meteors themselves are slow (about 23 km~s$^{-1}$) and fragile \citep[see][for a summary of observations]{borovicka2007}.

The first radar observations of the Draconids took place in the UK, during the 1946 outburst, with a military radar at a frequency of 60~MHz \citep{hey1947}. This radar had a power of 150 kW, and had a narrow, vertical beam which increased the gain. They observed a dramatic increase in meteor rates (from ~10/minute to 300/minute) on 1946 October 10 (peaking at solar longitude 197$^\circ$), lasting for about six hours. The echoes included many overdense (radiatively thick trails caused by larger meteoroids) and head echoes (scattering from the ionization around the meteoroid itself rather than the ionized train). The shower was also observed with the Jodrell Bank radar \citep{lovell1947}, frequency 72~MHz and transmitter power 150 kW, likewise with a narrow beam. It observed a peak of 200 echoes per minute at the same solar longitude.

\citet{davies1955} describe a radar survey using Jodrell Bank from 1947 to 1954 which specifically ran on the expected Draconid peak days, and found only one significant return of the shower in 1952, peaking at solar longitude 197$^\circ$. At this point the radar had been upgraded to have two independently steerable antennas \citep{aspinall1951}, which were pointed just north and south of due west. The time of maximum is somewhat uncertain, since the radiant was close to zenith, meaning that radar echoes were close to the horizon. 

A moderate outburst in 1985 was observed by \citet{simek1986} with the Springhill meteor radar in Ottawa, Canada, which had a peak power of 1.8~MW and ran at 32.5~MHz. The high power of the radar meant that it was not normally used for meteor rate studies, because of the manual analysis required for each echo, but a small portion of the data during the shower peak was analysed carefully. A correction was performed to correct the observed rates for the gain pattern of the antennas, which is the first step required to obtain fluxes, though formal collecting areas were not calculated. Overdense and underdense echoes were treated separately, and the sporadic background, based on observations in 1967, was subtracted. There was a peak at 9 UT 1985 October 8 (solar longitude 195.2$^\circ$), with a comparable second peak in the overdense (larger meteoroids) rates one hour later, though the scatter in number of echoes was large and the peak time therefore uncertain. 

\citet{simek1999} observed another Draconid outburst in 1998 using the Ond\v{r}ejov Observatory radar. This radar has a broad beam which was directed perpendicular to the radiant, and only underdense echoes were used for the analysis. The frequency of the radar is 37.5~MHz, and the transmitter power 20~kW. This was the first radar flux measurement of the Draconids, using the gain pattern of the antennas to determine the collecting area of the system. The authors found a peak flux of 0.162 meteoroids~km$^{-2}$~hr$^{-1}$ at a solar longitude of 195.1$^\circ$.

A 1999 Draconid outburst at radar magnitudes (faint meteors) was post-predicted by \citet{egal2019}, and a search of data from the early CMOR radar, moved to Alert, Canada, for the 1999 Leonids, found observations of the outburst. More details are given in the following sections.

The 2005 outburst was observed with CMOR \citep{campbell2006}, and is described in more detail below; similarly, the 2011 and 2012 outbursts were likewise observed with CMOR \citep{ye2013,ye2014}. The 2011 and 2012 outbursts were also seen by the Shigaraki middle and upper atmosphere (MU) radar in Japan \citep{kero2012,fujiwara2016}. This radar, at 46.5~MHz and 1~MW power, detects primarily head echoes, and does not therefore have to use statistical methods to calculate shower activity, since the orbits of meteors can be determined. Only 13 Draconid meteors were observed in 2011 because the radiant was below the horizon during the peak of the shower at solar longitude 195$^\circ$. The radiant was also low in 2012, but 57 Draconids were detected, and correcting for the radiant elevation showed a peak activity at a solar longitude of 195.6$^\circ$, in reasonable agreement with the CMOR results. The MU study also confirmed that the 2012 outburst was very rich in faint meteors, which explains why it was not a major outburst in optical observations. Simulations by \citet{Kastinen2017} support the enhanced delivery of smaller masses to the Earth in 2012 compared to 2011.

The mass index of the shower, $s$, describes how mass is distributed in the stream by particle size. An $s$ of 2 indicates that there is equal mass in each size bin; $s>2$ indicates more mass in small particles, and $s<2$ means there is more mass in large particles. Typical shower mass indices fall between 1.70 and 2.0, with sporadic meteors normally having an index in the range 2.0 to 2.3.  There is a wide scatter in mass index measurements for the Draconids. In 1985, \citet{simek1994} found that for underdense echoes, the mass index was 2.06, and 2.11 for larger, overdense meteors. In 1998, visual observations showed a scattered $s$ between 1.75 and 2.36 \citep{arlt1998}, while slightly fainter video observations gave an $s$ of 1.81 \citep{watanabe1999}. In 2005, video observations gave an $s$ of 1.87 at the peak and 1.78 for the full distribution \citep{koten2007}, while the CMOR data gave $s$=2.0 \citep{campbell2006}. In 2011, the CMOR observations \citep{ye2013} gave a mass index of 1.75, while those of 2012 \citep{ye2014} gave an $s$ of 1.88. Visual, video and photographic observations \citep{kac2015} gave an overall $s$ in 2011 of 2.0.  Video observations in 2018 by \citet{vida2020} showed a rapidly changing mass index, from 1.74 in the last part of the peak (the first part of the peak being obscured by twilight), to 2.32 immediately after. This agrees reasonably with the video observations of \citet{koten2020}, which found an $s$ which varied from 2.04 to 1.70, with the lowest value at the strongest part of the shower.

The current work examines radar observations of the flux of the most recent Draconid outbursts (1999 to 2018), using radars from Canada and Europe.

\section{Observations}

Six Skyimet meteor radars were used to calculate Draconid fluxes, five in Europe and one (with three frequencies) in Canada. All of the radars used a single transmitter antenna (one per frequency for the Canadian radar), and an array of five receiving antennas which allows the direction angles of each echo to be determined \citep{jones1998}. The Andenes radar in Norway (69.3$^\circ$N, 16.0$^\circ$E) operates at 32.55 MHz, with a power of 30 kW at the time of the observations used in this study and  crossed dipole antennas with circular polarization. The Andenes meteor radar has previously been used for meteor flux calculations \citep{Stober_2013_Geminids}. The Juliusruh radar in Germany (54.6$^\circ$N, 13.4$^\circ$E) has a very similar system at 32.55 MHz, at a power of 15 kW: the second Juliusruh system at 53.5 MHz was not used for this study, since the initial trail radius effect is worse for higher frequencies. The radar was transmitting and receiving circular polarization during the 2018 observations.  More information on these radars is available in \citet{wilhelm2019}. The Collm radar in Germany (51.3$^\circ$N, 13$^\circ$E) operates at 36.2 MHz and has a power of 15 kW (6 kW until 2015) and had a linear polarization transmit and receive antenna configuration. In 2016 the antennas were upgraded to circular polarization. It was assumed all the European Skyimet radars had a calibration coefficient (to convert measured echo amplitude to received power, used to calculate a limiting magnitude) of $4.9\times 10^{-21}$ W/du$^2$, where du is the digital units in which amplitude is measured, as measured in \citet{stober2011} for the Collm system. The last two European radars are Esrange at Kiruna in Sweden (67.9$^\circ$N, 21.1$^\circ$E), at a frequency of 32.5 MHz and a transmitter power of 6 kW, and Sodankyl\"{a} in Finland (64.7$^\circ$N, 26.6$^\circ$E), at 36.9 MHz and 15 kW. Esrange was running in 1999 during the Draconids, but on the day of the outburst there were issues with the interferometry, making it impossible to pick out the radiants of echoes. Both Esrange and Sokankyl\"{a} were running correctly during the 2011 and 2012 outbursts, but neither gathered useful data in 2018. 

The CMOR radar operates at three frequencies: 17.45, 29.85 and 38.15 MHz. The 29 MHz system has orbital capabilities using five remote receivers, but these orbits were not used for the current study, since the rate of meteors with orbits depends on environmental factors and is not easily corrected for flux studies. The power on all three systems is continuously monitored: the 29 MHz system was operating at 15.5 kW during the 2011, 2012 and 2018 outbursts. The 17 and 38 MHz systems have powers of about 6 kW. The 17 MHz system suffers significant terrestrial interference because of its frequency, and did not gather useful data during the 2005 Draconid outburst: it was also not operational between February 2011 and May 2013 and therefore missed the 2011 and 2012 outbursts, so only 2018 data is available. The CMOR radar has been in operation since 1999, but has only had stable power and receiver calibrations since 2002; in general, no fluxes are calculated prior to this. The exception is observations made from Alert, Canada (82.455$^\circ$N, 62.497$^\circ$W), when the 29 and 38 MHz systems were deployed in October and November 1999 to observe the 1999 Leonid outburst; the 29 MHz system was running during the 1999 Draconids. The power and receiver calibrations were not monitored continuously during these 1999 observations, as they are in the radar's final location near Tavistock, Ontario (43.264$^\circ$N, 80.772$^\circ$W), but the power on the 29 MHz system was measured on 1999 October 12 at 3.46 kW, and that is the power used to calculate the fluxes for 1999. The receivers were not calibrated, so the first calibration after they were returned to Ontario in January 2000 was used. During the 2011 outburst, the echo line at the predicted maximum was very close to the horizon, so the 38 MHz system antennas were rotated by 90 degrees, so that the maximum sensitivity of the antennas was pointed toward the center of the echo line. 

Details of all the radars used in this study are given in Table~\ref{table:radars}.

\begin{table*}
\caption{Details of radars used to observe Draconid outbursts }
\label{table:radars}
\centering
\begin{tabular}{lccccc}
\hline
%\vline
Radar & Frequency & Lat & Long & Power & Years observed\\
 & (MHz) & $^\circ$N & $^\circ$E & (kW) \\
\hline
%\vline
Alert & 29.85 & 82.455 & 297.50 & 3.46 & 1999\\
CMOR & 17.45 & 43.264 & 279.228 & 15 & 2018\\
CMOR & 29.85 & 43.264 & 279.228 & 15 & 2005, 2011, 2012, 2018\\
CMOR & 38.15 & 43.264 & 279.228 & 6 & 2005, 2011, 2012, 2018\\
Collm & 36.2 & 51.3 & 13.0 & 6/15 & 2011\\
Juliusruh & 32.55 & 54.6 & 13.4 & 15 & 2018\\
Andenes & 32.55 & 69.3 & 16.0 & 30 & 2011, 2012, 2018\\
Kiruna & 32.5 & 67.9 & 21.1 & 6 & 2011, 2012\\
Sodankyl\"{a} & 36.9 & 67.4 & 26.6 & 15 & 2011, 2012\\
\hline
\end{tabular}
\end{table*}

\section{Flux Calculations}

All the Skyimet radars used in this study, with the exception of the 29 MHz CMOR system, are single-station radars, meaning that while the location of each meteor echo can be determined through interferometry, the radiant of the meteor is unknown. Since all underdense meteor echoes are specular (at right angles to the trail), the number of echoes coming from a particular radiant can be determined statistically. To find the flux of a shower, all echoes within 3 degrees of being perpendicular to the radiant (lying on a great circle 90 degrees from the radiant) are counted as potential shower meteors. To remove sporadic meteors contaminating the shower echo line, activity on the echo line five days before and after the shower activity period are measured, and a baseline sporadic activity during the shower is interpolated for the points in between. This is subtracted from the shower activity. The fraction of echoes subtracted varied from about 5\% during the peaks of 1999, 2005 2011 and 2018 to less than 2\% at the peak in 2012. 

To calculate flux, the area of this echo strip is calculated at 15-minute intervals. The short intervals are needed for brief outbursts like the Draconids, and justified because the echo count is high enough that the noise is tolerable. Each small section of the echo strip is weighted for the range of the echoes (there is a $\propto R^{-3}$ decrease in sensitivity with range for specular echoes). In addition, Faraday rotation (the rotation of the polarization of the radar beam as it travels through free electrons in the ionosphere in the presence of the Earth's magnetic field) is calculated for linearly polarized antennas: it does not affect circularly polarized antennas. Faraday rotation is especially important at higher latitudes, where the magnetic field is nearly vertical. The gain of the transmitter and receiver antennas must also be taken into account in calculating the collecting area. In all cases, the mass index $s$  is used to correct the area to a constant sensitivity; less sensitive parts of the echo strip will miss meteor echoes, and the distribution of masses is needed to determine what fraction are missed. 

Finally, the calculated flux must be corrected for the initial radius effect. We are using the work of \citet{jones2005a}, which used $10^4$ sporadic meteors simultaneously observed on CMOR's 29 and 38 MHz systems. There is substantial scatter in the observed attenuation in the ratio of amplitudes at the two wavelengths, which indicates that the initial radius effect is different for individual meteoroids. If the Draconids have substantially larger initial radii than average sporadic meteors (because of a lateral spread in fragments, for example, which may be both more numerous and released earlier than typical given the Draconids fragile nature), their initial radius correction is underestimated, and both the 38 MHz and (to a lesser extent) the 29 MHz fluxes will need additional correction to higher values.

The fluxes calculated here are determined using the code from \citet{campbell2006}, with a few small changes. A correction has been added for solar activity \citep{campbell2019}, generally of order 10\% or less, increasing measured fluxes during times of high solar activity as measured with the F10.7 flux. The geometric limits of the collecting area have also been changed; instead of 22 degrees above the horizon, the limit is a specific sensitivity, which varies for antenna gains which are not radially symmetric. In practice, the new limit is very close to the old, only a few degrees less or more, but it makes a difference if the echo line is very close to the horizon, as is frequently the case for the Draconid shower. Finally, a correction has been added to compensate for noise on the receiver antennas, and the limiting magnitude is adjusted according to the measured noise of the system for the day. This was done only for the 29 and 38 MHz systems, where the noise characteristics are well studied.

\section{Results}

\begin{figure}
	\includegraphics[width=3.3in]{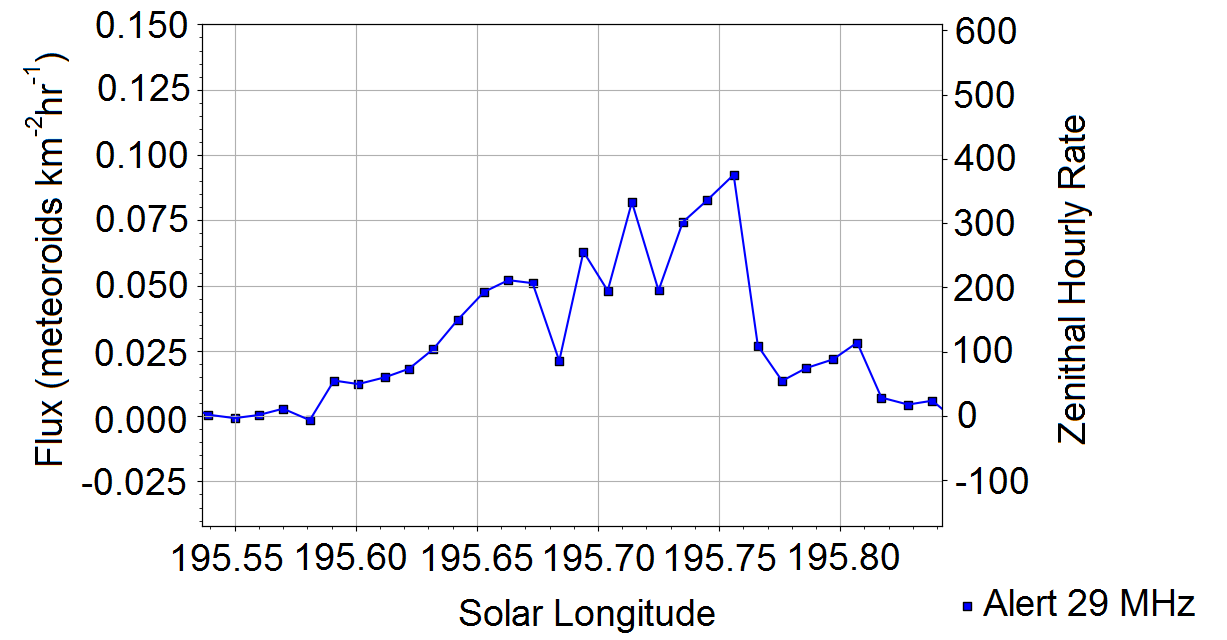}
    \caption{Draconid fluxes at 15-minute intervals for 1999. This interval is 1999 October 9 from approximately 7 to 15 UT. }
    \label{fig:flux1999}
\end{figure}

The fluxes from the 1999 outburst have already been published in \citet{egal2019}, but we have recalculated the fluxes with the updated flux code, shown in Fig.~\ref{fig:flux1999}. Since no receiver calibration was available, the absolute values of these fluxes compared to the other CMOR fluxes are uncertain. For all the flux plots, ZHR is given on the right axis for comparison. An $s$ of 1.8 was used for all years, and no attempt was made to change the value of $s$ during each shower. We have converted the flux into ZHR (Zenithal Hourly Rate), commonly used in visual observations, for comparison with other sources, on the right axis. This is the number of meteors an ideal observer would see in one hour if the radiant were at the zenith and the observer's limiting magnitude were +6.5.

\begin{figure}
	\includegraphics[width=3.3in]{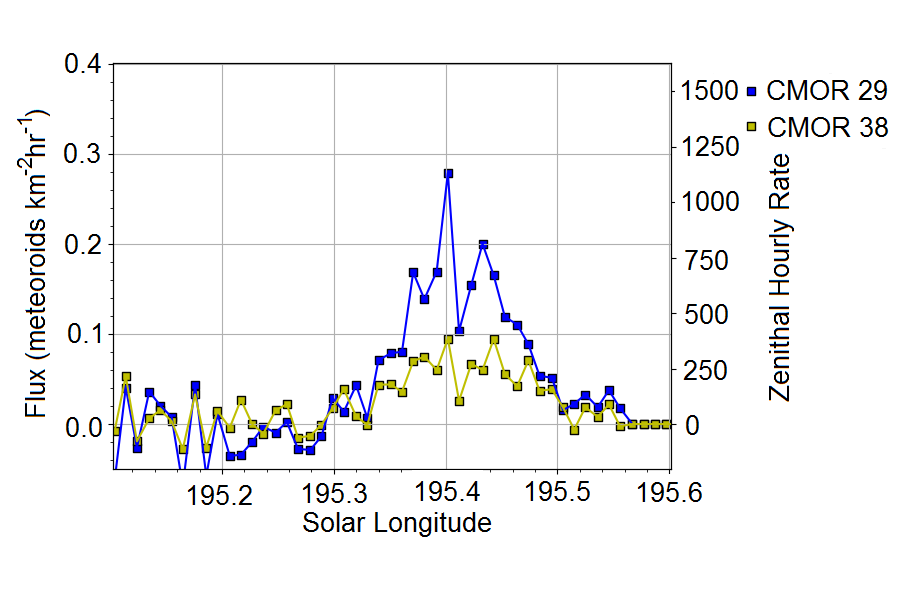}
    \caption{Draconid fluxes at 15-minute intervals for 2005 October 8, from approximately 13.5 to 18.5 UT. }
    \label{fig:flux2005}
\end{figure}

Likewise, we have recomputed the 2005 Draconid fluxes for the 29 and 38 MHz CMOR systems (Fig.~\ref{fig:flux2005}). Notice the flux on the 29 MHz is about twice the flux on 38 MHz. 

\begin{figure}
	\includegraphics[width=3.3in]{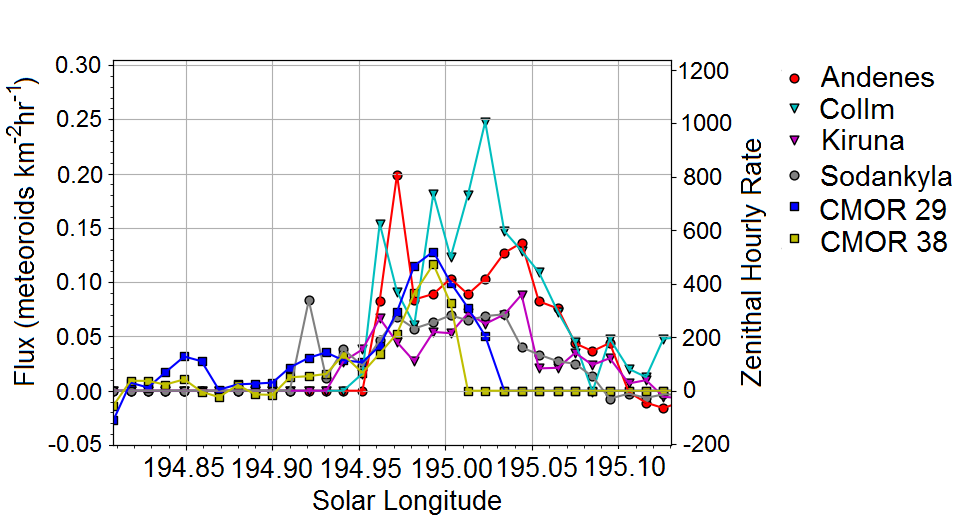}
    \caption{Draconid fluxes at 15-minute intervals for 2011 October 8, from approximately 17 to 23 UT. }
    \label{fig:flux2011}
\end{figure}

\begin{figure}
	\includegraphics[width=3.3in]{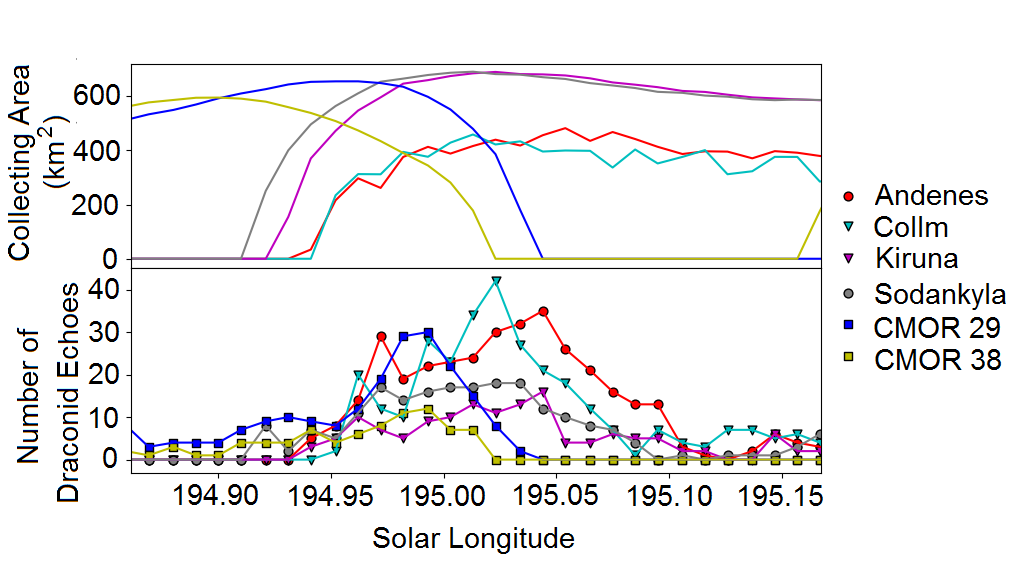}
    \caption{Draconid collecting areas (top) and raw echo line numbers (bottom) at 15-minute intervals for 2011 }
    \label{fig:CAnum2011}
\end{figure}

\begin{figure}
	\includegraphics[width=3.3in]{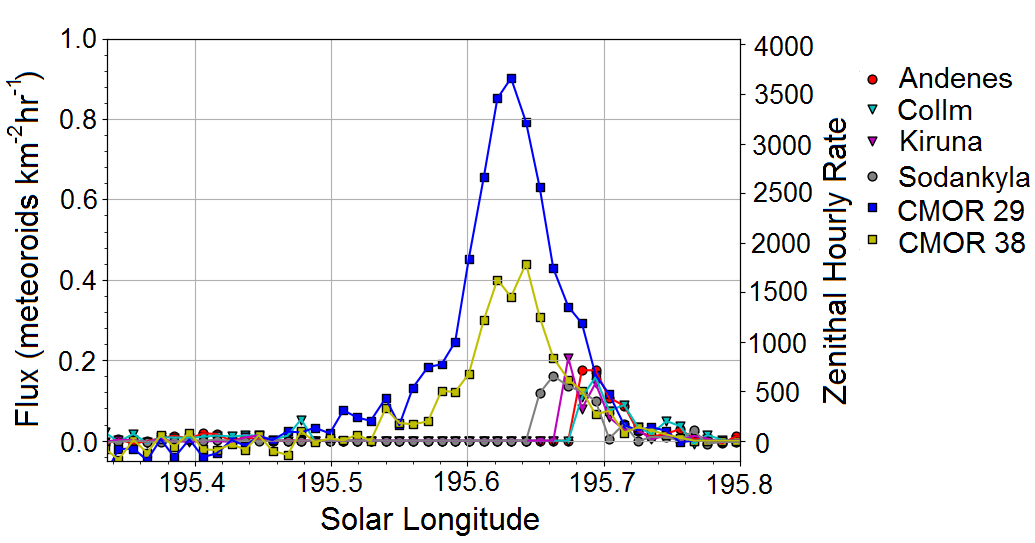}
    \caption{Draconid fluxes at 15-minute intervals for 2012 October 8, from approximately 13.5 to 20 UT. }
    \label{fig:flux2012}
\end{figure}

\begin{figure}
	\includegraphics[width=3.3in]{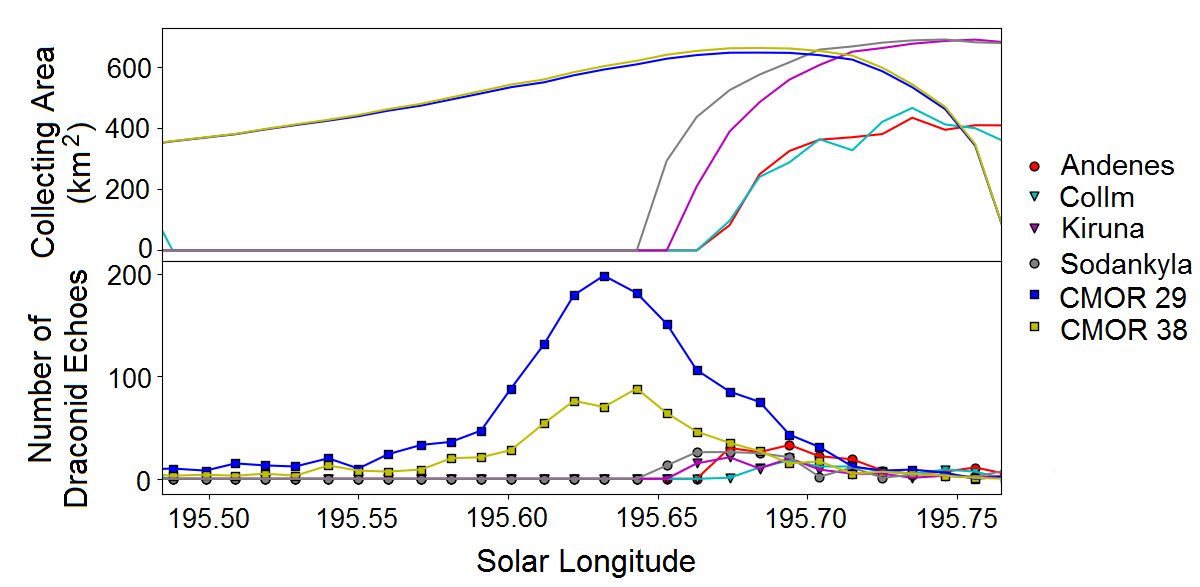}
    \caption{Draconid collecting areas (top) and raw echo line numbers (bottom) at 15-minute intervals for 2012 October 8/9, from approximately 19 to 05 UT. }
    \label{fig:CAnum2012}
\end{figure}

In 2011 and 2012, there are fluxes from six radars (Figs.~\ref{fig:flux2011} and \ref{fig:flux2012}), but in each case some or all of the radars did not observe the whole shower. In 2011, the beginning of the shower was observed by CMOR in North America, and the end was observed by the European radars. Fig.~\ref{fig:CAnum2011} shows the collecting area of each of the systems. The two CMOR systems have different collecting areas in this year because of the rotation of the 38 MHz system antennas, which shifted the time during which the radiant was too low to be observed. The European radars observed only the end of the strong 2012 return; Fig.~\ref{fig:CAnum2012} shows the raw number of echoes on the echo line and the collecting area for each radar in 2012. In 2011 and 2012 Collm is strongly affected by Faraday rotation due to the echo line passing through the magnetic field direction; Andenes, Kiruna and Sodankyl\"{a} had antennas with circular polarization and were therefore not affected. 

\begin{figure}
	\includegraphics[width=3.3in]{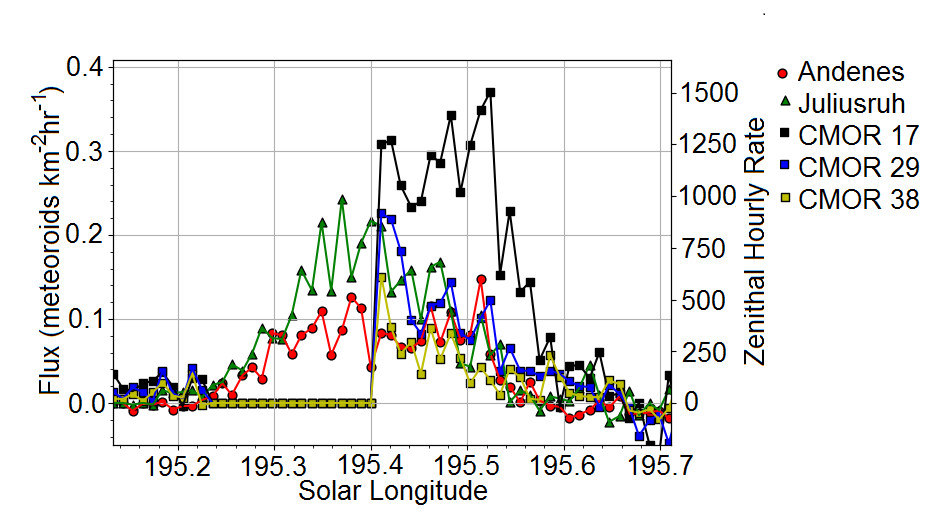}
    \caption{Draconid fluxes at 15-minute intervals for 2018 }
    \label{fig:flux2018}
\end{figure}

\begin{figure}
	\includegraphics[width=3.3in]{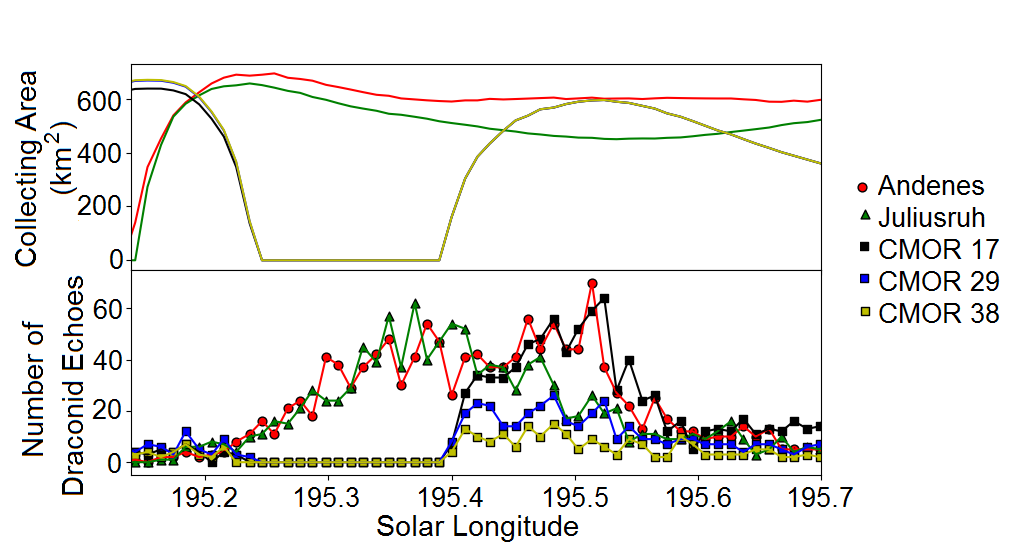}
    \caption{Draconid collecting areas (top) and raw echo line numbers (bottom) at 15-minute intervals for 2018 }
    \label{fig:CAnum2018}
\end{figure}

The fluxes for 2018 are fully recorded on Andenes and Juliusruh: geometry prevented CMOR from observing until midway through the outburst. The progression in flux from 38 to 17 MHz is clearly seen in Fig.~\ref{fig:flux2018}, confirming that the difference between 29 and 38 MHz is a frequency-dependent effect. Fig.~\ref{fig:CAnum2018} shows the collecting area for the five systems which observed the 2018 outburst: the three CMOR radars have nearly identical collecting areas, except that 17 MHz is affected by Faraday rotation in the afternoon and the collecting area is slightly lower at that time (it did not affect shower observations in 2018). 

\begin{figure*}
	\includegraphics[width=7in]{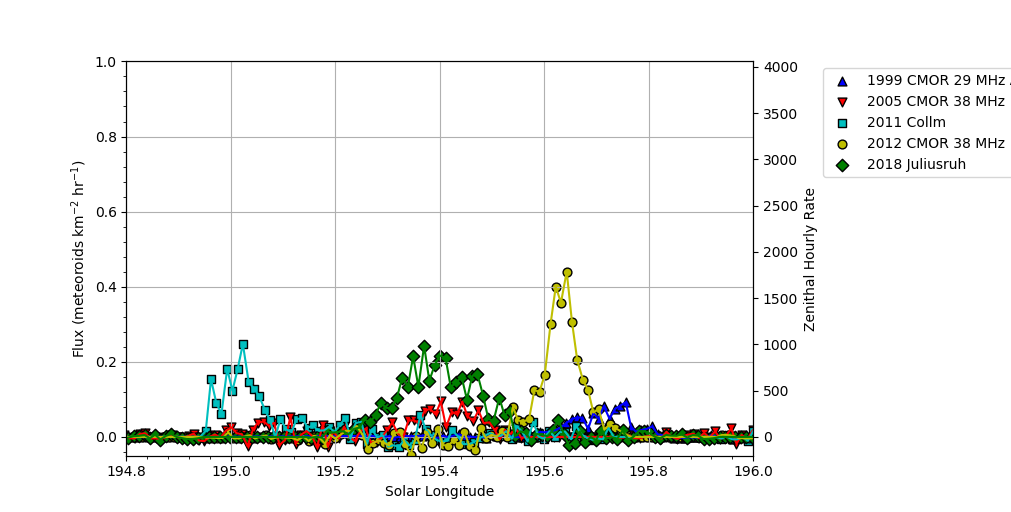}
    \caption{Draconid fluxes at 15-minute intervals for all years, selected radars }
    \label{fig:fluxallyears}
\end{figure*}

For clarity, all five of these outbursts are plotted on a single graph in Fig.~\ref{fig:fluxallyears}. For each year, the best data from a single radar was used: 38 MHz was used where possible for CMOR, since that radar has the best year-to-year calibration, but it only fully observed two of the outbursts. The absolute scales of the different radars may not be the same, but the timing of the showers can be clearly seen. The timing and range of peak ZHRs are shown in Table~\ref{table:outbursts}. The maximum solar longitude is an intermediate value reflecting all the radars which observed the outburst; those with fewer radars observing have more precise values, but all the values likely have uncertainties of the order of $\pm$0.1 degrees in solar longitude, as discussed below. The table also gives the times at which the outburst began and ended (note that the 2018 outburst began on October 8 and ended the following day, UT), and the range of peak ZHRs observed by all the radars. Again, a smaller range of observed ZHRs reflects only a lack of multiple observations, not higher precision observations.

\begin{table*}
\caption{Characteristics of Draconid outbursts, 1999 to 2018. ZHR range refers to the range of peak ZHRs measured by all radar systems observing in that year. Note that in some years the peak solar longitude is different on different systems, by up to 0.1 degrees.}
\label{table:outbursts}
\centering
\begin{tabular}{lccccc}
\hline
%\vline
Year & Peak Solar longitude ($^\circ$) & Date & Begin time (UT) & End time (UT) & ZHR range\\
\hline
%\vline
1999 & 195.75 & Oct 9 & 7 & 15 & 380\\
2005 & 195.4 & Oct 8 & 13.5 & 18.5 & 400 - 1150\\
2011 & 195 & Oct 8 & 17 & 23 & 300 - 1000\\
2012 & 195.63 & Oct 8 & 13.5 & 20 & 1700 - 3600\\
2018 & 195.4 & Oct 8/9 & 19 & 5 & 600 - 1500\\
\hline
\end{tabular}
\end{table*}

\section{Discussion}

The absolute fluxes on different radars agree to within a factor of two, which is encouraging given that many of the radars do not have regular calibration data collected. At the 15-minute intervals chosen for this study, different radars have different times for the maximum flux. The noise is significant at these short intervals, even when there are tens of echoes in each bin. It is obvious in 2011 and 2018 in particular, when there are many systems observing the peak, that different systems have different local maxima; this indicates that the timing of the maximum cannot be determined to better than a couple of hours precision. 

In most years, the flux measured on the 29 MHz system was approximately twice that measured on the 38 MHz system. This is consistent with an undercorrection for the initial radius effect, which affects shorter wavelengths (higher frequencies) more strongly. If, for example, Draconid meteors produce trails with larger radii due to greater fragmentation than average meteors, the standard initial radius correction would not be sufficient. This is supported by the three frequency observations in 2018, where the 17 MHz system (with a much longer wavelength) shows the highest flux. In 2011, the 29 MHz fluxes were only 40\% higher than the 38 MHz fluxes: whether that is because the properties of the meteoroids in that year were different or because of an uncompensated systematic offset in the 29 MHz system is not clear. The 38 MHz system has a much steadier baseline because of the lack of system upgrades \citep[see][for details]{campbell2019}, while the 29 MHz system shows much greater excursions from year to year.  
The European radars have fluxes which generally agree more closely with the 29 MHz system, although since simultaneous power and receiver calibrations are not available for these systems it is not possible, for example, to use their data to characterize the initial radius correction factor. All of the European Skyimet systems have frequencies between 29 and 38 MHz. 

Although the fluxes from different years are not necessarily comparable due to differences in calibrations, it is interesting that the integrated fluxes (flux-time products) visible in Fig.~\ref{fig:fluxallyears} agree reasonably with the simulations of \citet{Kastinen2017} and \citet{egal2019}.

The Draconid outbursts are all short, lasting less than half a day in all cases, and often less than six hours. The data here show the importance of having radars with a spread in longitudes to characterize the shape and magnitude of the activity: even with continuous operation, no single location could have observed all of the outbursts since 1999 from beginning to end. For locations further north than +35$^\circ$, the Draconid radiant is circumpolar, but when the radiant approaches the zenith, the echo strip at right angles is nearly at the horizon, where the antenna gain is low and ranges are ambiguous. This means that the times at which specular meteor radars can observe the shower are restricted to radiant elevations under $\approx$70$^\circ$. For high-latitude showers like the Draconids, this results in a substantial blind period each day, even while the shower is above the horizon. Optical methods are not subject to the same difficulty, but most optical observations missed the peak of the 2012 outburst, which occurred during the day in North America and Europe. High-power, large aperture radars observing primarily head echoes do not have this issue, but are mostly (with the exception of EISCAT and MAARSY in northern Scandinavia) located at low northern latitudes or further south, where the radiant is below the horizon at least some of the time \citep[e.g.][]{kero2012}. These high power radars are not normally run in a mode compatible with meteor observations, and tend to see fewer shower meteors, since sporadics increasingly dominate at smaller sizes \citep[see ][]{kero2019,SCHULT_2017_MAARSY_sporadics,SCHULT_2018_MAARSY_shower}.

\section{Conclusions}

Five recent outbursts of the Draconids, observed with radars, have been compared. There is broad agreement in the shape of the outbursts from system to system, even in very different geographical locations. There is more uncertainty in the level of activity, stressing the need for good calibrations and bias corrections. For the Draconid shower in particular, it is evident that an improvement is needed in the initial radius correction factor, which apparently needs to be larger. 

These observations also show the importance of observations with a spread in geographical longitude for short-duration outbursts like the Draconid showers. In general, no single location will provide continuous coverage with a specular meteor radar.

\section*{Acknowledgements}

Funding  for  this  work  was  provided  through  NASA  cooperative agreement 80NSSSC18M0046 and the Natural Sciences and Engineering Research Council of Canada (Grant no. RGPIN-2018-05474). The Esrange meteor radar operation, maintenance and data collection is provided by Esrange Space Center of Swedish Space Corporation (SSC). GS is a member of the Oeschger Center for Climate Change Research. The Andenes and Juliusruh meteor radar data was collected under the grant STO 1053/1-1 of the Deutsche Forschungsgemeinschaft (DFG). We thank Jorge L. Chau and R. Latteck for their support of the AHEAD project.

\section*{Data Availability}

Data available on request.
%%%%%%%%%%%%%%%%%%%%%%%%%%%%%%%%%%%%%%%%%%%%%%%%%%



\bibliographystyle{mnras}
\bibliography{dracradar} % if your bibtex file is called example.bib

\begin{thebibliography}{}
\makeatletter
\relax
\def\mn@urlcharsother{\let\do\@makeother \do\$\do\&\do\#\do\^\do\_\do\%\do\~}
\def\mn@doi{\begingroup\mn@urlcharsother \@ifnextchar [ {\mn@doi@}
  {\mn@doi@[]}}
\def\mn@doi@[#1]#2{\def\@tempa{#1}\ifx\@tempa\@empty \href
  {http://dx.doi.org/#2} {doi:#2}\else \href {http://dx.doi.org/#2} {#1}\fi
  \endgroup}
\def\mn@eprint#1#2{\mn@eprint@#1:#2::\@nil}
\def\mn@eprint@arXiv#1{\href {http://arxiv.org/abs/#1} {{\tt arXiv:#1}}}
\def\mn@eprint@dblp#1{\href {http://dblp.uni-trier.de/rec/bibtex/#1.xml}
  {dblp:#1}}
\def\mn@eprint@#1:#2:#3:#4\@nil{\def\@tempa {#1}\def\@tempb {#2}\def\@tempc
  {#3}\ifx \@tempc \@empty \let \@tempc \@tempb \let \@tempb \@tempa \fi \ifx
  \@tempb \@empty \def\@tempb {arXiv}\fi \@ifundefined
  {mn@eprint@\@tempb}{\@tempb:\@tempc}{\expandafter \expandafter \csname
  mn@eprint@\@tempb\endcsname \expandafter{\@tempc}}}

\bibitem[\protect\citeauthoryear{{Arlt}}{{Arlt}}{1998}]{arlt1998}
{Arlt} R.,  1998, WGN, 26, 256

\bibitem[\protect\citeauthoryear{{Aspinall}, {Clegg}  \& {Hawkins}}{{Aspinall}
  et~al.}{1951}]{aspinall1951}
{Aspinall} A.,  {Clegg} A.,   {Hawkins} G.,  1951, Phil. Mag., 42, 504

\bibitem[\protect\citeauthoryear{{Borovi{\v{c}}ka}, {Spurn{\'y}}  \&
  {Koten}}{{Borovi{\v{c}}ka} et~al.}{2007}]{borovicka2007}
{Borovi{\v{c}}ka} J.,  {Spurn{\'y}} P.,   {Koten} P.,  2007, \mn@doi [\aap]
  {10.1051/0004-6361:20078131}, \href
  {https://ui.adsabs.harvard.edu/abs/2007A&A...473..661B} {473, 661}

\bibitem[\protect\citeauthoryear{{Campbell-Brown}}{{Campbell-Brown}}{2019}]{campbell2019}
{Campbell-Brown} M.~D.,  2019, \mn@doi [\mnras] {10.1093/mnras/stz697}, \href
  {https://ui.adsabs.harvard.edu/abs/2019MNRAS.485.4446C} {485, 4446}

\bibitem[\protect\citeauthoryear{{Campbell-Brown}, {Vaubaillon}, {Brown},
  {Weryk}  \& {Arlt}}{{Campbell-Brown} et~al.}{2006}]{campbell2006}
{Campbell-Brown} M.,  {Vaubaillon} J.,  {Brown} P.,  {Weryk} R.~J.,   {Arlt}
  R.,  2006, \mn@doi [\aap] {10.1051/0004-6361:20054588}, \href
  {https://ui.adsabs.harvard.edu/abs/2006A&A...451..339C} {451, 339}

\bibitem[\protect\citeauthoryear{Davies \& Lovell}{Davies \&
  Lovell}{1955}]{davies1955}
Davies J.~G.,  Lovell A.,  1955, Monthly Notices of the Royal {\ldots}, 115, 23

\bibitem[\protect\citeauthoryear{{Egal}, {Wiegert}, {Brown}, {Moser},
  {Campbell-Brown}, {Moorhead}, {Ehlert}  \& {Moticska}}{{Egal}
  et~al.}{2019}]{egal2019}
{Egal} A.,  {Wiegert} P.,  {Brown} P.~G.,  {Moser} D.~E.,  {Campbell-Brown} M.,
   {Moorhead} A.,  {Ehlert} S.,   {Moticska} N.,  2019, \mn@doi [\icarus]
  {10.1016/j.icarus.2019.04.021}, \href
  {https://ui.adsabs.harvard.edu/abs/2019Icar..330..123E} {330, 123}

\bibitem[\protect\citeauthoryear{{Fujiwara}, {Kero}, {Abo}, {Szasz}  \&
  {Nakamura}}{{Fujiwara} et~al.}{2016}]{fujiwara2016}
{Fujiwara} Y.,  {Kero} J.,  {Abo} M.,  {Szasz} C.,   {Nakamura} T.,  2016,
  \mn@doi [\mnras] {10.1093/mnras/stv2492}, \href
  {https://ui.adsabs.harvard.edu/abs/2016MNRAS.455.3273F} {455, 3273}

\bibitem[\protect\citeauthoryear{{Hey}, {Parsons}  \& {Stewart}}{{Hey}
  et~al.}{1947}]{hey1947}
{Hey} J.~S.,  {Parsons} S.~J.,   {Stewart} G.~S.,  1947, \mn@doi [\mnras]
  {10.1093/mnras/107.2.176}, \href
  {https://ui.adsabs.harvard.edu/abs/1947MNRAS.107..176H} {107, 176}

\bibitem[\protect\citeauthoryear{{Jones} \& {Campbell-Brown}}{{Jones} \&
  {Campbell-Brown}}{2005}]{jones2005a}
{Jones} J.,  {Campbell-Brown} M.,  2005, \mn@doi [\mnras]
  {10.1111/j.1365-2966.2005.08972.x}, \href
  {https://ui.adsabs.harvard.edu/abs/2005MNRAS.359.1131J} {359, 1131}

\bibitem[\protect\citeauthoryear{{Jones}, {Webster}  \& {Hocking}}{{Jones}
  et~al.}{1998}]{jones1998}
{Jones} J.,  {Webster} A.~R.,   {Hocking} W.~K.,  1998, \mn@doi [Radio Science]
  {10.1029/97RS03050}, \href
  {https://ui.adsabs.harvard.edu/abs/1998RaSc...33...55J} {33, 55}

\bibitem[\protect\citeauthoryear{{Kac}}{{Kac}}{2015}]{kac2015}
{Kac} J.,  2015, WGN, Journal of the International Meteor Organization, \href
  {https://ui.adsabs.harvard.edu/abs/2015JIMO...43...75K} {43, 75}

\bibitem[\protect\citeauthoryear{{Kastinen} \& {Kero}}{{Kastinen} \&
  {Kero}}{2017}]{Kastinen2017}
{Kastinen} D.,  {Kero} J.,  2017, \mn@doi [\planss]
  {10.1016/j.pss.2017.03.007}, \href
  {https://ui.adsabs.harvard.edu/abs/2017P&SS..143...53K} {143, 53}

\bibitem[\protect\citeauthoryear{{Kero}, {Fujiwara}, {Abo}, {Szasz}  \&
  {Nakamura}}{{Kero} et~al.}{2012}]{kero2012}
{Kero} J.,  {Fujiwara} Y.,  {Abo} M.,  {Szasz} C.,   {Nakamura} T.,  2012,
  \mn@doi [\mnras] {10.1111/j.1365-2966.2012.21255.x}, \href
  {https://ui.adsabs.harvard.edu/abs/2012MNRAS.424.1799K} {424, 1799}

\bibitem[\protect\citeauthoryear{{Kero}, {Campbell-Brown}, {Stober}, {Chau},
  {Mathews}  \& {Pellinen-Wannberg}}{{Kero} et~al.}{2019}]{kero2019}
{Kero} J.,  {Campbell-Brown} M.~D.,  {Stober} G.,  {Chau} J.~L.,  {Mathews}
  J.~D.,   {Pellinen-Wannberg} A.,  2019, {Radar Observations of Meteors}.
p.~65

\bibitem[\protect\citeauthoryear{Koten, Borovi{\v{c}}ka, Spurn{\'{y}}  \&
  Stork}{Koten et~al.}{2007}]{koten2007}
Koten P.,  Borovi{\v{c}}ka J.,  Spurn{\'{y}} P.,   Stork R.,  2007, \mn@doi
  [Astronomy and Astrophysics] {10.1051/0004-6361}, 466, 729

\bibitem[\protect\citeauthoryear{{Koten} et~al.,}{{Koten}
  et~al.}{2020}]{koten2020}
{Koten} P.,  et~al., 2020, \mn@doi [\planss] {10.1016/j.pss.2020.104871}, \href
  {https://ui.adsabs.harvard.edu/abs/2020P&SS..18404871K} {184, 104871}

\bibitem[\protect\citeauthoryear{{Lovell}, {Banwell}  \& {Clegg}}{{Lovell}
  et~al.}{1947}]{lovell1947}
{Lovell} A.~C.~B.,  {Banwell} C.~J.,   {Clegg} J.~A.,  1947, \mn@doi [\mnras]
  {10.1093/mnras/107.2.164}, \href
  {https://ui.adsabs.harvard.edu/abs/1947MNRAS.107..164L} {107, 164}

\bibitem[\protect\citeauthoryear{Schult, Stober, Janches  \& Chau}{Schult
  et~al.}{2017}]{SCHULT_2017_MAARSY_sporadics}
Schult C.,  Stober G.,  Janches D.,   Chau J.~L.,  2017, \mn@doi [Icarus]
  {https://doi.org/10.1016/j.icarus.2017.06.019}, 297, 1

\bibitem[\protect\citeauthoryear{Schult, Brown, Pokorný, Stober  \&
  Chau}{Schult et~al.}{2018}]{SCHULT_2018_MAARSY_shower}
Schult C.,  Brown P.,  Pokorný P.,  Stober G.,   Chau J.~L.,  2018, \mn@doi
  [Icarus] {https://doi.org/10.1016/j.icarus.2018.02.032}, 309, 177

\bibitem[\protect\citeauthoryear{{\v{S}imek}}{{\v{S}imek}}{1986}]{simek1986}
{\v{S}imek} M.,  1986, Bulletin of the Astronomical Institutes of
  Czechoslovakia, \href {https://ui.adsabs.harvard.edu/abs/1986BAICz..37..246S}
  {37, 246}

\bibitem[\protect\citeauthoryear{{\v{S}imek}}{{\v{S}imek}}{1994}]{simek1994}
{\v{S}imek} M.,  1994, \aap, \href
  {https://ui.adsabs.harvard.edu/abs/1994A&A...284..276S} {284, 276}

\bibitem[\protect\citeauthoryear{{\v{S}imek} \& {Pecina}}{{\v{S}imek} \&
  {Pecina}}{1999}]{simek1999}
{\v{S}imek} M.,  {Pecina} P.,  1999, \aap, \href
  {https://ui.adsabs.harvard.edu/abs/1999A&A...343L..94S} {343, L94}

\bibitem[\protect\citeauthoryear{{Stober}, {Singer}  \& {Jacobi}}{{Stober}
  et~al.}{2011}]{stober2011}
{Stober} G.,  {Singer} W.,   {Jacobi} C.,  2011, \mn@doi [Journal of
  Atmospheric and Solar-Terrestrial Physics] {10.1016/j.jastp.2010.07.018},
  \href {https://ui.adsabs.harvard.edu/abs/2011JASTP..73.1069S} {73, 1069}

\bibitem[\protect\citeauthoryear{Stober, Schult, Baumann, Latteck  \&
  Rapp}{Stober et~al.}{2013}]{Stober_2013_Geminids}
Stober G.,  Schult C.,  Baumann C.,  Latteck R.,   Rapp M.,  2013, \mn@doi
  [Annales Geophysicae] {10.5194/angeo-31-473-2013}, 31, 473

\bibitem[\protect\citeauthoryear{{Vida}, {Campbell-Brown}, {Brown}, {Egal}  \&
  {Mazur}}{{Vida} et~al.}{2020}]{vida2020}
{Vida} D.,  {Campbell-Brown} M.,  {Brown} P.~G.,  {Egal} A.,   {Mazur} M.~J.,
  2020, \mn@doi [\aap] {10.1051/0004-6361/201937296}, \href
  {https://ui.adsabs.harvard.edu/abs/2020A&A...635A.153V} {635, A153}

\bibitem[\protect\citeauthoryear{{Watanabe}, {Abe}, {Takanashi}, {Hashimoto},
  {Iiyama}, {Ishibashi}, {Morishige}  \& {Yokogawa}}{{Watanabe}
  et~al.}{1999}]{watanabe1999}
{Watanabe} J.-i.,  {Abe} S.,  {Takanashi} M.,  {Hashimoto} T.,  {Iiyama} O.,
  {Ishibashi} Y.,  {Morishige} K.,   {Yokogawa} S.,  1999, \mn@doi [\grl]
  {10.1029/1999GL900195}, \href
  {https://ui.adsabs.harvard.edu/abs/1999GeoRL..26.1117W} {26, 1117}

\bibitem[\protect\citeauthoryear{{Wilhelm}, {Stober}  \& {Brown}}{{Wilhelm}
  et~al.}{2019}]{wilhelm2019}
{Wilhelm} S.,  {Stober} G.,   {Brown} P.,  2019, \mn@doi [Annales Geophysicae]
  {10.5194/angeo-37-851-2019}, \href
  {https://ui.adsabs.harvard.edu/abs/2019AnGeo..37..851W} {37, 851}

\bibitem[\protect\citeauthoryear{{Ye}, {Brown}, {Campbell-Brown}  \&
  {Weryk}}{{Ye} et~al.}{2013}]{ye2013}
{Ye} Q.,  {Brown} P.~G.,  {Campbell-Brown} M.~D.,   {Weryk} R.~J.,  2013,
  \mn@doi [\mnras] {10.1093/mnras/stt1605}, \href
  {https://ui.adsabs.harvard.edu/abs/2013MNRAS.436..675Y} {436, 675}

\bibitem[\protect\citeauthoryear{{Ye}, {Wiegert}, {Brown}, {Campbell-Brown}  \&
  {Weryk}}{{Ye} et~al.}{2014}]{ye2014}
{Ye} Q.,  {Wiegert} P.~A.,  {Brown} P.~G.,  {Campbell-Brown} M.~D.,   {Weryk}
  R.~J.,  2014, \mn@doi [\mnras] {10.1093/mnras/stt2178}, \href
  {https://ui.adsabs.harvard.edu/abs/2014MNRAS.437.3812Y} {437, 3812}

\makeatother
\end{thebibliography}

%\begin{thebibliography}{99}
%\end{thebibliography}

%%%%%%%%%%%%%%%%%%%%%%%%%%%%%%%%%%%%%%%%%%%%%%%%%%

%%%%%%%%%%%%%%%%% APPENDICES %%%%%%%%%%%%%%%%%%%%%

%\appendix



%%%%%%%%%%%%%%%%%%%%%%%%%%%%%%%%%%%%%%%%%%%%%%%%%%

% Don't change these lines
\bsp	% typesetting comment
\label{lastpage}
\end{document}